\documentclass [prc,aps,showpacs,twocolumn]{revtex4}
\usepackage{graphicx}
\begin{document}

\newcommand{\pd}[2]{\frac{\partial #1}{\partial #2}}
\newcommand{\pdline}[2]{{\partial #1}/{\partial #2}}
\newcommand{\be}{\begin{equation}}
\newcommand{\ee}[1]{ \label{#1}  \end{equation}}

\newcommand{\adag}{a^{\dag}}
\newcommand{\atil}{\tilde{a}}
\def\frp*1{${*1\over2}^+$}
\def\frm*1{${*1\over2}^-$}
\def\g{\noindent}
\def\mev{\hbox{\MeV}}
\def\kev{\hbox{\keV}}
\def\lambdabar{{\mathchar'26\mkern-9mu\lambda}}
\def\lambdabarrr{{^-\mkern-12mu\lambda} }
% \large{
% \draft

\title{R\'enyi statistics in equilibrium statistical mechanics}

\author{A.S.~Parvan$^{a,b}$  and  T.S.~Bir\'{o}$^{c,}$}

\affiliation{$^{a}$Joint Institute for Nuclear Research, BLTP, 141980 Dubna, Russian Federation}

\affiliation{$^{b}$Institute of Applied Physics, Moldova Academy of Sciences, MD-2028 Chisinau, Republic of Moldova}

\affiliation{$^{c}$KFKI Research Institute for Particle and Nuclear Physics, H-1525 Budapest, P.O.Box 49, Hungary}

%\email{parvan@theor.jinr.ru}
%\thanks{(Alexandru Parvan)}

%\date{\today}

\begin{abstract}
The R\'enyi statistics in the canonical and microcanonical ensembles is examined in the general case and in particular for the ideal gas. In the microcanonical ensemble the R\'enyi statistics is equivalent with the Boltzmann-Gibbs statistics. By the exact analytical results for the ideal gas, it is shown that in the canonical ensemble in the thermodynamic limit the R\'enyi statistics is also equivalent with the Boltzmann-Gibbs statistics. Furthermore it satisfies the requirements of the equilibrium thermodynamics, i.e. the thermodynamical potential of the statistical ensemble is a homogeneous function of degree $1$ of its extensive variables of state. We conclude that the R\'enyi statistics duplicates the thermodynamical relations stemming from the Boltzmann-Gibbs statistics in the thermodynamical limit.
\end{abstract}


\maketitle

%%%%%%%%%%%%%%%%%%%%%%%%%%%%%%%% SECTION 1 %%%%%%%%%%%%%%%%%%%%%%%%
\section{Introduction}
The general aim of this paper is to establish the connection of the R\'enyi statistics in the canonical and microcanonical ensembles with the equilibrium thermodynamics and to compare it with the Boltzmann-Gibbs statistical mechanics. The R\'{e}nyi statistics is defined by the R\'enyi statistical entropy~\cite{Renyi,Wehrl} with respect to the usual norm equation for probabilities and the usual  expectation values for dynamical variables~\cite{Lenzi00,ParvBiro}. It differs from the Boltzmann-Gibbs statistics only in the definition of the statistical entropy. The main advantage of the R\'enyi statistics is the power-law distribution function instead of the exponential distribution function of the Gibbs statistics. Therefore the R\'enyi entropy~\cite{Renyi,Wehrl} has been applied in several situations, for example, fractal and multifractal systems~\cite{Beck,Jizba}, high energy particle production~\cite{Kropivnitskaya}, nuclear physics~\cite{Nagy} and black holes~\cite{Bialas}. However in these fields the Tsallis statistics is more significant   due to the nonextensive nature of the Tsallis statistical entropy~\cite{Tsal88,Tsal98}.

The R\'enyi statistics as the theory of the equilibrium statistical mechanics is defined at the equilibrium in the maximum entropy point. Therefore it must satisfy all requirements of the equilibrium thermodynamics and must be in agreement with the theory of probability, and some general physical principles. The R\'enyi statistics is thermodynamically self-consistent whenever it satisfies in the thermodynamic limit the zero, first, second, and third laws of thermodynamics, the principle of additivity, the fundamental equation of thermodynamics, the Gibbs-Duhem relation, the differential equation for thermodynamical  potential, the Legendre transform and the Euler theorem. It can be proved that the zeroth law of thermodynamics, the principle of additivity, the Euler theorem, and the Gibbs-Duhem relation are valid if the thermodynamical potential of the statistical ensemble is a homogeneous function of first degree of the extensive variables of state~\cite{Parv1,Parv2}. For finite systems the thermodynamic potential is an inhomogeneous function. Therefore finite systems are not thermodynamically self-consistent. In the thermodynamical limit the thermodynamical potential of the statistical ensemble is a homogeneous function of the first degree. Therefore for infinite systems the connection of the statistical mechanics with the equilibrium thermodynamics is provided. The equilibrium thermodynamics is a macroscopic phenomenological theory which is defined in the thermodynamic limit, while boundary effects can be neglected~\cite{Kvasn}.

The R\'enyi statistical entropy is an extensive function if it is a multivariate function of the set of probabilities. Therefore the R\'enyi statistics was considered thermodynamically self-consistent without any further verification. Otherwise the Tsallis statistics was thoroughly investigated. The studies done by many authors fail to answer the question whether the zeroth law of thermodynamics and the principle of additivity for the Tsallis statistics are valid. These unsuccessful attempts are based on the following physical concepts: the physical temperature~\cite{Abe0}, the entropic parameter $q$ as universal constant~\cite{Abe1,Wang2}, the temperature dependent nonextensive Hamiltonian~\cite{Wang1}, the transformations from the Tsallis entropy to the R\'enyi (Gibbs) one~\cite{Abe3,Johal,Fa,ParvBiro} and the inconsequent thermodynamic limit~\cite{Abe2}. In~\cite{Parv1,Parv2}, the author proved that the zeroth law of thermodynamics in the thermodynamic limit for the Tsallis statistics is satisfied only if the parameter $z=1/(q-1)$ is an extensive variable of state. Notice that the consistent thermodynamic limit for the Tsallis statistics was firstly proposed in the particular case by Botet et al. in~\cite{Botet1,Botet2}.

In our previous paper~\cite{ParvBiro}, we obtained the main relations for the R\'enyi statistics in the canonical ensemble in the general case and for the ideal gas. It was shown that the R\'enyi statistics in the microcanonical ensemble is equivalent with the usual Boltzmann-Gibbs. In the present paper we will show that the R\'enyi statistics in the canonical and microcanonical ensembles in the thermodynamical limit is thermodynamically self-consistent whenever it coincides with the usual Boltzmann-Gibbs statistics.

The authors in~\cite{Oikon,Bagci} erroneously claimed that the R\'enyi statistics has the exponential distribution function instead of the power-law probability distribution function. In~\cite{Oikon} this follows as a wrong interpretation for the distribution function of the microcanonical ensemble and in~\cite{Bagci} the exponential stationary distribution function was obtained on the basis of a generalized entropy maximization procedure, distinct from the usual R\'enyi one.

This paper is organized as follows. In the second section we briefly outline the canonical ensemble for the R\'{e}nyi statistics in the general form and for the ideal gas and define the zeroth law of thermodynamics. The same results for the microcanonical ensemble are given in the third section of the present paper.

%%%%%%%%%%%%%%%%%%%%%%%%%%%%%%%% SECTION 2 %%%%%%%%%%%%%%%%%%%%%%%%
\section{Canonical ensemble $(T,V,z,N)$}
\subsection{General formalism}
The equilibrium statistical mechanics is said to be the R\'{e}nyi statistics if the equilibrium phase space distribution function satisfies conditions imposed by the R\'{e}nyi's statistical entropy~\cite{Renyi,Wehrl} with respect to the norm equation for a phase space distribution function and an expectation value for a dynamical variable $A$~\cite{Lenzi00,ParvBiro}
\begin{eqnarray}\label{1}
    S &=& k\frac{\ln\left(\int d\Gamma \varrho^{q} \right)}{1-q} = -kz \ln\left(\int d\Gamma \varrho^{1+\frac{1}{z}} \right), \\ \label{2}
   1 &=& \int d\Gamma \varrho, \\ \label{3}
 \langle A\rangle &=& \int d\Gamma \varrho A,
\end{eqnarray}
where $\varrho$ is the phase space distribution function, $d\Gamma=dxdp$ is an infinitesimal element of phase space, $k$ is the Boltzmann constant, $q\in \mathbf{R}$ is the real parameter, $q\in [0,\infty]$, and $z=\frac{1}{q-1}$ is a variable of state.

Consider the canonical ensemble of the classical dynamical systems of $N$ particles at a constant temperature $T$, in a volume $V$, and with a thermodynamic coordinate $z$, in a thermal contact with a heat bath. To express the equilibrium phase space distribution function from the macroscopic variables of state, we consider the thermodynamical method based on the fundamental equation of thermodynamics~\cite{Parv1,Parv2} instead of the Jaynes principle~\cite{Jaynes}. The fundamental equation of thermodynamics at the fixed values of $T,V,z,N$ can be written as
\begin{equation}\label{4}
     (T dS-d\langle H\rangle)_{T,V,z,N} = 0.
\end{equation}
Using Eqs.~(\ref{1}) and (\ref{3}), we obtain
\begin{eqnarray}\nonumber
   \int  d\Gamma  \left[T \frac{\delta S}{\delta \varrho}- \frac{\delta \langle H\rangle}{\delta \varrho}\right] d\varrho + T\frac{\partial S}{\partial z}dz &-& \\ \label{5}
     - \int  d\Gamma   \frac{\delta\langle H\rangle}{\delta H}\ dH &=& 0,
\end{eqnarray}
where the expectation value of the Hamiltonian $H$ is
\begin{equation}\label{6}
\langle H\rangle = \int d\Gamma \varrho H.
\end{equation}
Since in the canonical ensemble differentials $dH=dz=d\varrho=0$, we obtain
\begin{equation}\label{7}
    T\frac{\delta S}{\delta \varrho}-
    \frac{\delta \langle H\rangle}{\delta \varrho}=\alpha_{1},
    \;\; \frac{\partial S}{\partial z} = \alpha_{2}, \;\;
\frac{\delta\langle H\rangle}{\delta H}  = \alpha_{3},
\end{equation}
where $\alpha_{1},\alpha_{2}$ and $\alpha_{3}$ are certain constants. Substituting Eqs.~(\ref{1}) and (\ref{6}) in the first equation (\ref{7}), we get
\begin{equation}\label{8}
    -kT (z+1) \varrho^{1/z} e^{\frac{S}{kz}} - H = \alpha_{1}.
\end{equation}
Averaging Eq.~(\ref{8}) and eliminating the constant $\alpha_{1}$, we obtain the phase space distribution function
\begin{equation}\label{9}
\varrho = \frac{1}{Z_{R}} \left[1-\frac{H}{kT(z+1)+\langle H\rangle}\right]^{z},
\end{equation}
where
\begin{eqnarray}\label{10}
Z_{R} &=& \int d\Gamma \left[1-\frac{H}{kT(z+1)+\langle H\rangle}\right]^{z}, \\ \label{11}
 \langle H\rangle &=&  \frac{1}{Z_{R}} \int d\Gamma H \left[1-\frac{H}{kT(z+1)+\langle H\rangle}\right]^{z}.
\end{eqnarray}
The norm equations (\ref{10}), (\ref{11}) are derived from Eqs.~(\ref{2}) and (\ref{6}). Therefore the phase space distribution function is determined by two norm functions of the variables of state,
$Z_{R}=Z_{R}(T,V,z,N)$ and $\langle H\rangle=\langle H\rangle(T,V,z,N)$, respectively. They are two solutions of the system of equations (\ref{10}), (\ref{11}). The thermodynamical system is totally determined by the free energy. It is the Legendre transform of energy with respect to the entropy of the system and it can be written as
\begin{eqnarray}\label{12}
    F &\equiv& \langle H\rangle-TS = \\ \nonumber
    &=& -kT \ln\left[Z_{R}  e^{-\frac{\langle H\rangle}{kT}} \left(1+\frac{\langle H\rangle}{kT(z+1)}\right)^{z}
   \right],
\end{eqnarray}
where
\begin{equation}\label{13}
    S=k \ln\left[Z_{R} \left(1+\frac{\langle H\rangle}{kT(z+1)}\right)^{z} \right].
\end{equation}
The expectation value of a dynamical variable $A$ can be written as
\begin{equation}\label{14}
 \langle A\rangle = \frac{1}{Z_{R}} \int d\Gamma A \left[1-\frac{H}{kT(z+1)+\langle H\rangle}\right]^{z}.
\end{equation}

The fundamental equation of thermodynamics for R\'{e}nyi statistics is derived at constant values of the variables of state $(T,V,z,N)$ following the procedure given in Ref.~\cite{Parv2}, we obtain
\begin{equation}\label{15}
     T dS= d\langle H\rangle + pdV +X dz -\mu dN,
\end{equation}
where
\begin{eqnarray}\label{16}
  p &=& \int d\Gamma \varrho  \left(-\frac{\partial H}{\partial V}  \right)_{T,z,N} , \\ \label{17}
  \mu &=& \int d\Gamma \varrho \left(\frac{\partial H}{\partial N} \right)_{T,V,z} , \\ \label{18}
    X &=& kT\left[\frac{S}{kz}+e^{\frac{S}{kz}} \int d\Gamma \varrho^{1+\frac{1}{z}}\ln\varrho^{\frac{1}{z}}
    \right].
\end{eqnarray}
Here the property of the Hamilton function, $(\partial H/\partial T)_{V,z,N}=(\partial H/\partial z)_{T,V,N}=0$, is used.

The thermodynamic potential $F$ of the canonical ensemble $(T,V,z,N)$, i.e., the Helmholtz free energy (\ref{12}), satisfies the differential relation which can be written as
\begin{equation}\label{20}
    dF= -S dT - p dV - X dz + \mu dN,
\end{equation}
where
\begin{eqnarray}\label{21}
  S &=& -\left(\frac{\partial F}{\partial T}\right)_{V,z,N}, \;\;\;
  p = -\left(\frac{\partial F}{\partial V}\right)_{T,z,N}, \\ \label{22}
  X &=& -\left(\frac{\partial F}{\partial z}\right)_{T,V,N}, \;\;\;
  \mu = \left(\frac{\partial F}{\partial N}\right)_{T,V,z}.
\end{eqnarray}
The free energy is the thermodynamic potential and it determines the ensemble averages of the canonical ensemble.

The fundamental equation of thermodynamics (\ref{15}) provides the first and second laws of thermodynamics
\begin{equation}\label{23}
    \delta Q=TdS, \;\; \delta Q= d\langle H\rangle + pdV +X dz -\mu
    dN,
\end{equation}
where $\delta Q$ is a heat transfer by the system to the environment during a quasistatic transition of the system from one equilibrium state to another nearby one. The heat capacity, $\delta Q =C dT$, in the canonical ensemble at the fixed values of $T,V,z,N$ can be written as
\begin{eqnarray}\label{24} \nonumber
    C_{VzN} &=& \left(\frac{\delta Q}{dT}\right)_{V,z,N}=
    T\left(\frac{\partial S}{\partial T}\right)_{V,z,N}= \\
    &=&\left(\frac{\partial \langle H\rangle}{\partial T}\right)_{V,z,N}=
    -T\left(\frac{\partial^{2} F}{\partial T^{2}}\right)_{V,z,N}.
\end{eqnarray}
Note that the thermodynamical relations for quantum R\'{e}nyi statistics are the same as for the classical one given above.

Let us prove the zeroth law of thermodynamics for the R\'enyi statistics in the canonical ensemble. Divide the system into two parts, $H = H_{1}+H_{2}$, under the conditions
\begin{eqnarray}\label{25}
  T &=& T_{1}=T_{2}, \;  N = N_{1}+N_{2}, \; V=V_{1}+V_{2},  \\ \label{26}
 z &=& z_{1}+z_{2}  \qquad \mbox{z-extensive},  \\ \label{26a}
z &=& z_{1}=z_{2} \qquad \mbox{z-intensive}.
\end{eqnarray}
The heat transfer $\delta Q$ is additive. Therefore, from the second law of thermodynamics (\ref{23}) it follows that the zeroth law, $T = T_{1}=T_{2}$, is valid whenever the total differential of entropy $dS$ is additive
\begin{eqnarray}\label{1b}
   \delta Q &=& \delta Q_{1} + \delta Q_{2}= T_{1}dS_{1}+ T_{2}dS_{2}= TdS, \\ \label{2b}
  dS &=& dS_{1}+ dS_{2}, \qquad T = T_{1}=T_{2}.
\end{eqnarray}
Let us show that Eq.~(\ref{2b}) is valid only if the entropy $S$ is an additive function
\begin{equation}\label{3b}
    S(T,V,z,N) = S_{1}(T_{1},V_{1},z_{1},N_{1}) + S_{2}(T_{2},V_{2},z_{2},N_{2}).
\end{equation}
Differentiating Eq.~(\ref{3b}) with respect to variables of state $T,V$ è $N$, we obtain
\begin{eqnarray}\label{4b}
  \frac{\partial S}{\partial T} &=& \frac{\partial S_{1}}{\partial T_{1}} +   \frac{\partial S_{2}}{\partial T_{2}},
  \qquad T = T_{1}=T_{2},\\ \label{5b}
   \frac{\partial S}{\partial V} &=& \frac{\partial S_{1}}{\partial V_{1}} =   \frac{\partial S_{2}}{\partial V_{2}},
   \qquad V=V_{1}+V_{2},\\ \label{6b}
   \frac{\partial S}{\partial N} &=& \frac{\partial S_{1}}{\partial N_{1}} =   \frac{\partial S_{2}}{\partial N_{2}}, \qquad N = N_{1}+N_{2}.
\end{eqnarray}
Differentiating Eq.~(\ref{3b}) with respect to variable of state $z$, we have two possibilities
\begin{eqnarray}\label{7b}
  \frac{\partial S}{\partial z} &=& \frac{\partial S_{1}}{\partial z_{1}} =   \frac{\partial S_{2}}{\partial z_{2}},
  \qquad z = z_{1}+z_{2},\\ \label{8b}
   \frac{\partial S}{\partial z} &=& \frac{\partial S_{1}}{\partial z_{1}} +   \frac{\partial S_{2}}{\partial z_{2}},
   \qquad z=z_{1}=z_{2}.
\end{eqnarray}
Then from Eqs.~(\ref{4b})--(\ref{8b}) and the differentials of Eqs.~(\ref{25})--(\ref{26a}), we get
\begin{equation}\label{9b}
    dS =dS_{1}+dS_{2}.
\end{equation}
Therefore the zeroth law of thermodynamics, $T = T_{1}=T_{2}$, is satisfied whenever the entropy is additive, $S=S_{1}+S_{2}$.

Let us show that the additivity of entropy (\ref{3b}) corresponds to the homogeneity of the function of entropy. Let $S$ be the homogeneous function of first degree of its arguments $(T,V,z,N)$. Then for $z$ extensive or $z$ intensive, we have, respectively,
\begin{eqnarray}\label{10b}
  S(T,\lambda V,\lambda z,\lambda N) &=& \lambda S(T,V,z,N), \quad \lambda=\frac{1}{N}, \;\;\;\;  \\ \label{11b}
  S(T,\lambda V,z,\lambda N)  &=& \lambda S(T,V,z,N).
\end{eqnarray}
Divide the system into two parts under the conditions (\ref{25})--(\ref{26a}). From Eq.~(\ref{25}) for $N$, we have
\begin{equation}\label{12b}
    \frac{1}{\lambda} =\frac{1}{\lambda_{1}}+\frac{1}{\lambda_{2}}.
\end{equation}
Then the intensive specific variables correspond to the additive (extensive) variables
\begin{eqnarray}\label{13b}
\lambda V=\lambda_{1} V_{1}=\lambda_{2} V_{2}, \quad \lambda z &=&\lambda_{1} z_{1} =\lambda_{2} z_{2},
\end{eqnarray}
where $\lambda_{1}=1/N_{1}$ and $\lambda_{2}=1/N_{2}$, while $\lambda=1/N$. Here $z$ is an extensive variable. Under the conditions of Eqs.~(\ref{3b}) and (\ref{10b})--(\ref{13b}), it follows that the specific entropy $s=\lambda S$ is an intensive function \begin{eqnarray}\label{15b} \nonumber
   \lambda S(T,V,z,N) &=& \lambda_{1} S_{1}(T_{1},V_{1},z_{1},N_{1})= \\
    &=& \lambda_{2} S_{2}(T_{2},V_{2},z_{2},N_{2}).
\end{eqnarray}
Conversely, from Eqs.~(\ref{15b}) and (\ref{10b})--(\ref{13b}), we obtain the additivity of entropy (\ref{3b}). Therefore the additivity of entropy (\ref{3b}) is a consequence of its homogeneity (\ref{10b}), (\ref{11b}) (extensivity). Thus the zeroth law of thermodynamics for R\'enyi statistics in the canonical ensemble is valid whenever the entropy (or the thermodynamic potential) is a homogeneous function of first degree of {\em all} extensive variables of state~\cite{Parv1,Parv2}.

If the entropy is a homogeneous function of degree one (\ref{10b}), (\ref{11b}), then $S$ satisfies the Euler theorem and the Gibbs-Duhem relation. Differentiating Eqs.~(\ref{10b}), (\ref{11b}) with respect to $\lambda$, and putting $\lambda=1$, we obtain the well-known Euler theorem for the homogeneous functions:
\begin{eqnarray} \label{16b}
  V\frac{\partial S} {\partial V} +
    z \frac{\partial S} {\partial z}  +
    N \frac{\partial S} {\partial N}  &=& S  \quad \mbox{z-extensive},\\ \label{17b}
   V \frac{\partial S} {\partial V}  +
      N \frac{\partial S}{\partial N} &=& S \quad \mbox{z-intensive}.
\end{eqnarray}
Using the Legendre transform (\ref{12}), the homogeneity property of energy $\langle H\rangle$, and the thermodynamic relations (\ref{21}), (\ref{22}), we get the Euler theorem~\cite{Prigogine}:
\begin{eqnarray} \label{18b}
 T S &=& \langle H\rangle+p V +Xz -\mu N  \qquad \mbox{z-extensive}, \\ \label{19b}
 T S &=& \langle H\rangle+p V -\mu N  \qquad \mbox{z-intensive}.
\end{eqnarray}
Applying the differential operator with respect to the ensemble variables $(T,V,z,N)$ on Eqs.~(\ref{18b}), (\ref{19b}) and using the fundamental equation of thermodynamics (\ref{15}), we obtain the Gibbs-Duhem relation~\cite{Prigogine}
\begin{eqnarray} \label{20b}
 SdT &=& Vdp +zdX -Nd\mu  \qquad \mbox{z-extensive}, \\ \label{21b}
 SdT &=& Vdp -Xdz -Nd\mu   \qquad \mbox{z-intensive}.
\end{eqnarray}
Therefore in the canonical ensemble the Euler theorem results from the homogeneity property of the entropy (\ref{10b}), (\ref{11b}), the energy $\langle H\rangle$, and the thermodynamic potential.
For $z$ being extensive the Euler theorem  (\ref{18b}) contains the term $Xz$ and it is consistent with the fundamental equation of thermodynamics (\ref{15}) and with the Gibbs-Duhem relation (\ref{20b}). However for $z$ being intensive the Gibbs-Duhem relation (\ref{21b}) is violated, and the Euler theorem (\ref{19b}) is not consistent  with the fundamental equation of thermodynamics (\ref{15}).
Thus the variable of state $z$ can only be extensive.

Dividing the system into two parts under the conditions (\ref{25},\ref{26a}),
i.e. assuming that $z$ is intensive,
we obtain that the phase space distribution function (\ref{9})--(\ref{11}) does not factorize and the R\'{e}nyi entropy (\ref{1}), (\ref{13}) is nonadditive
\begin{equation}\label{27}
 S\neq S_{1}+S_{2}, \qquad  \varrho\neq\varrho_{1}\varrho_{2}.
\end{equation}
Thus, the zeroth law of thermodynamics and the principle of additivity for the R\'{e}nyi statistics in the canonical ensemble are violated. In the general case for finite values of the variables of state $V,z,N$ the thermodynamical potential (\ref{12}) of the canonical ensemble is not a homogenous function of first order. Therefore the Euler theorem for finite systems in the canonical ensemble is not satisfied,
\begin{eqnarray}\label{19}
 T S &\neq& \langle H\rangle+p V +Xz -\mu N  \quad \mbox{z-extensive}, \\ \label{19bb}
 T S &\neq& \langle H\rangle+p V -\mu N  \quad \mbox{z-intensive}.
\end{eqnarray}
The proof of inequalities (\ref{27})--(\ref{19bb}) is given below for the ideal gas.

In the general case, the R\'{e}nyi entropy (\ref{1}), (\ref{13}) and the thermodynamical potential (\ref{12}) of the canonical ensemble is not a homogeneous function. But it can become homogeneous in the thermodynamic limit. A limiting statistical procedure is called the thermodynamic limit if the following conditions hold: $(i)$ $N\rightarrow\infty$, $V\rightarrow\infty$, $z\rightarrow\infty$ and $v=V/N =\mathrm{const}$, $\tilde{z}=z/N =\mathrm{const}$ for $z$ extensive; $(ii)$ $N\rightarrow\infty$, $V\rightarrow\infty$ and $v=V/N =\mathrm{const}$, $z =\mathrm{const}$ for $z$ intensive. This means that the functions of variables of state can be expand in power series with the small parameter $1/N$ $(N\gg 1)$ at large finite values of the variables $V,z$ for $z$ extensive and large values of $V$ for $z$ intensive. Then the extensive function $\mathcal{A}$ is said to be the series $(\alpha> 0)$
\begin{eqnarray} \label{22b} \nonumber
   \mathcal{A}(T,V,z,N) &=& N[a(T,v,\tilde{z}) + O(N^{-\alpha})] \\
    &\stackrel{as} =&  N a(T,v,\tilde{z})  \qquad \mbox{z-extensive}, \\
    \label{23b} \nonumber
   \mathcal{A}(T,V,z,N) &=& N[a(T,v,z) + O(N^{-\alpha})] \\
    &\stackrel{as} =&  N a(T,v,z)  \qquad \mbox{z-intensive},
\end{eqnarray}
and the intensive function $\phi$ is called the series
\begin{eqnarray} \label{24b} \nonumber
\phi(T,V,z,N) &=& \phi(T,v,\tilde{z}) + O(N^{-\alpha}) \\
      &\stackrel{as} =& \phi(T,v,\tilde{z})  \qquad \mbox{z-extensive}, \\
\label{25b} \nonumber
\phi(T,V,z,N) &=& \phi(T,v,z) + O(N^{-\alpha}) \\
      &\stackrel{as} =& \phi(T,v,z)  \qquad \mbox{z-intensive},
\end{eqnarray}
where $v=V/N$ is the specific volume, $\tilde{z}=z/N$ is the specific $z$, and $a =\mathcal{A} /N$ is the specific $\mathcal{A}$. Comparing (\ref{22b}), (\ref{23b}) and (\ref{10b}), (\ref{11b}), it follows that the entropy is a homogeneous function of degree $1$ only in the thermodynamic limit. Thus the zeroth law of thermodynamics, the Euler theorem and the Gibbs-Duhem relation can be satisfied only in the thermodynamic limit.

\subsection{Ideal gas in canonical ensemble}
The thermodynamic properties of the R\'{e}nyi statistics in the canonical ensemble $(T,V,z,N)$ can be investigated by the exact analytic functions of the variables of state for the classical nonrelativistic ideal gas. The main relations for the ideal gas were obtained by us in Ref.~\cite{ParvBiro}. Here we express them through the variable $z=1/(q-1)$ and study the thermodynamical properties of such a system. The Hamiltonian and the dimensionless infinitesimal element of phase space for the ideal gas of $N$ identical particles can be written as
\begin{equation}\label{28}
  d\Gamma = \frac{1}{N!(2\pi\hbar)^{3N}} \prod\limits_{i=1}^{N}
  d\vec{r}_{i} d\vec{p}_{i}, \qquad  H = \sum\limits_{i=1}^{N} \frac{\vec{p}_{i}^{2}}{2m}.
\end{equation}
The phase space distribution function (\ref{9}) and the norm functions $Z_{R}$ and $\langle H\rangle$ determined from the system of equations (\ref{10}) and (\ref{11}) take the form
\begin{equation}\label{30}
\varrho = \frac{1}{Z_{R}} \left[1-\frac{1}{kT(z+1)+\langle
H\rangle} \sum\limits_{i=1}^{N} \frac{\vec{p}_{i}^{2}}{2m}
\right]^{z},
\end{equation}
where
\begin{eqnarray}\label{31}
 Z_{R} &=& a_{z,N} \left( 1+\frac{3}{2} \frac{N}{z+1}\right)^{-z}  Z_{G},  \\
 \label{32}
 \langle H\rangle  &=& \langle H\rangle_{G} = \frac{3}{2} kTN
\end{eqnarray}
and
\begin{eqnarray}\label{33}
  a_{z,N} &=& \frac{\Gamma(z+1) \left( 1+\frac{3}{2}
  \frac{N}{z+1}\right)^{z+\frac{3}{2}N}}
  {(z+1)^{-\frac{3}{2}N}\Gamma(z+1+\frac{3}{2}N)}, \;\; z>0, \\ \label{34}
  a_{z,N} &=& \frac{\Gamma(-z-\frac{3}{2}N) \left( 1+\frac{3}{2}
  \frac{N}{z+1}\right)^{z+\frac{3}{2}N}}
  {(-z-1)^{-\frac{3}{2}N}\Gamma(-z)},  z<0,
\end{eqnarray}
The value of $z$ in Eq.~(\ref{34}) is restricted by the condition $z<-\frac{3}{2}N$. The Gibbs partition function in Eq.~(\ref{31}) is
\begin{equation}\label{35}
  Z_{G} = \frac{(gV)^{N}}{N!} \left(\frac{mkT}{2\pi
  \hbar^{2}}\right)^{\frac{3}{2}N}.
\end{equation}
In the following the subscript $G$ indicates the Gibbs statistics. In the Gibbs limit $z\to\pm\infty,N=const,V=const$ the R\'enyi statistics resembles the usual Boltzmann-Gibbs statistics. In this limit the phase space distribution function (\ref{30}) takes the form
\begin{equation}\label{58}
\varrho|_{z\to\pm\infty} = \varrho_{G} = \frac{1}{Z_{G}} \
e^{-\frac{1}{kT}\sum\limits_{i=1}^{N} \frac{\vec{p}_{i}^{2}}{2m}
},
\end{equation}
where the norm functions (\ref{31}) and (\ref{32}) are $Z_{R}|_{z\to\pm\infty} = Z_{G}$ and $\langle H\rangle = \langle H\rangle_{G}$. The function (\ref{33}), (\ref{34}) in the Gibbs limit is $a_{z,N}|_{z\to\pm\infty}=e^{3N/2}$.

The thermodynamic potential of the canonical ensemble for the classical ideal gas of the R\'{e}nyi statistics is found to be
\begin{equation}\label{36}
    F = F_{G}-kT \left(\ln a_{z,N}- \frac{3}{2}N\right),
\end{equation}
where $F_{G} = -kT \ln Z_{G}$ is the free energy of the Gibbs statistics. In the Gibbs limit the free energy (\ref{36}) is reduced to the form $F|_{z\to\pm\infty} = F_{G}$. From Eq.~(\ref{21}), we obtain the entropy $S$ and the pressure $p$ \begin{eqnarray}\label{37}
  S &=& S_{G} + k \left(\ln a_{z,N}- \frac{3}{2}N\right), \\
  \label{38}
  p &=& p_{G} = \frac{N}{V} kT = \frac{2}{3} \frac{\langle
  H\rangle}{V},
\end{eqnarray}
where
\begin{equation}\label{39}
    S_{G} = k\ln Z_{G} + \frac{3}{2} kN.
\end{equation}
In the Gibbs limit the entropy and pressure can be written as $S|_{z\to\pm\infty} = S_{G}$ and $p=p_{G}$. The chemical potential (\ref{22}) for the ideal gas of the R\'enyi statistics in the canonical ensemble can be written as
\begin{equation}\label{40}
    \mu = \mu_{G} - kT (A_{\mu}-\frac{3}{2}),
\end{equation}
where $A_{\mu}=\partial (\ln a_{z,N})/\partial N$ and
\begin{eqnarray}\label{41}
  A_{\mu} &=&
  \frac{3}{2}\left[\frac{z+\frac{3}{2}N}{z+1+\frac{3}{2}N}
  -\psi(z+1+\frac{3}{2}N)\right] + \nonumber \\
  &+& \frac{3}{2} \ln(z+1+\frac{3}{2}N), \qquad z>0, \\ \label{42}
  A_{\mu} &=&
  \frac{3}{2}\left[\frac{z+\frac{3}{2}N}{z+1+\frac{3}{2}N}
  -\psi(-z-\frac{3}{2}N)\right] + \nonumber \\
  &+& \frac{3}{2} \ln(-z-1-\frac{3}{2}N), \qquad z<-\frac{3}{2}N.
\end{eqnarray}
Here $\psi(z)$ is the psi-function and $\mu_{G}$ is the chemical potential of the Gibbs statistics. It takes the form
\begin{eqnarray} \label{43}
  \mu_{G} &=& \left(\frac{\partial F_{G}}{\partial
  N}\right)_{T,V} = kT\psi(N+1) - \nonumber \\
  &-& kT \ln\left[gV\left(\frac{mkT}{2\pi
  \hbar^{2}}\right)^{\frac{3}{2}}\right].
\end{eqnarray}
In the Gibbs limit the chemical potential (\ref{40}) is $\mu|_{z\to\pm\infty} = \mu_{G}$. The variable $X$ (\ref{22}) for the ideal gas of the R\'enyi statistics can be written as
\begin{equation}\label{44}
    X =  kT X_{a},
\end{equation}
where $X_{a}=\partial (\ln a_{z,N})/\partial z$ and
\begin{eqnarray}\label{45}
  X_{a} &=&
  \frac{3}{2}\frac{N}{z+1}\left[1-\frac{z+\frac{3}{2}N}{z+1+\frac{3}{2}N}\right] - \nonumber \\
  &-& \psi(z+1+\frac{3}{2}N) + \ln(1+\frac{3}{2}\frac{N}{z+1}) +
  \nonumber \\ &+&   \psi(z+1), \qquad z>0, \\ \label{46}
  X_{a} &=&
   \frac{3}{2}\frac{N}{z+1}\left[1-\frac{z+\frac{3}{2}N}{z+1+\frac{3}{2}N}\right] - \nonumber \\
  &-& \psi(-z-\frac{3}{2}N) + \ln(1+\frac{3}{2}\frac{N}{z+1}) +
  \nonumber \\ &+&   \psi(-z), \qquad z<-\frac{3}{2}N.
\end{eqnarray}
In the Gibbs limit the variable (\ref{44}) takes the form $X|_{z\to\pm\infty} = 0$. The heat capacity (\ref{24})  for the ideal gas of the R\'enyi statistics is now
\begin{equation}\label{47}
    C_{VzN} = C_{VN}^{G}=\frac{3}{2} kN.
\end{equation}
The heat capacity for R\'enyi statistics is the same as for Gibbs statistics.

The $N$-particle distribution function of the classical ideal gas in the canonical ensemble for R\'{e}nyi statistics  can be defined as
\begin{eqnarray}\label{48}
    f(\vec{p}_{1},\ldots,\vec{p}_{N}) &=&
    \frac{(gV)^{N}}{N!h^{3N}Z_{R}}  \\ &\times& \left[1-
    \frac{1}{kT(z+1)+\langle
H\rangle} \sum\limits_{i=1}^{N} \frac{\vec{p}_{i}^{2}}{2m}
\right]^{z}, \nonumber
\end{eqnarray}
which must be normalized to unity
\begin{equation}\label{49}
    \int d^{3}p_{1}\cdots d^{3}p_{N}
    f(\vec{p}_{1},\ldots,\vec{p}_{N})=1.
\end{equation}
In the Gibbs limit the $N$-particle distribution function (\ref{48}) is reduced to the usual Maxwell-Boltzmann form
\begin{equation}\label{63}
  f(\vec{p}_{1},\ldots,\vec{p}_{N})|_{z\to\pm\infty} =
    \frac{(gV)^{N}}{N!h^{3N}Z_{G}} \
    e^{-\sum\limits_{i=1}^{N} \frac{\vec{p}_{i}^{2}}{2mkT}}.
\end{equation}
Integrating Eq.~(\ref{48}) with respect to momenta $\vec{p}_{2},\ldots,\vec{p}_{N}$, we get the reduced single-particle distribution function
\begin{eqnarray}\label{50}
  f(\vec{p}) &=&
    \frac{d_{z,N}}{(2\pi mkT)^{3/2}} \\ &\times& \left[1-
    \frac{\vec{p}^{2}}{2mkT(z+1+\frac{3}{2}N)}
    \right]^{z+\frac{3}{2}(N-1)}, \nonumber
\end{eqnarray}
where
\begin{eqnarray}\label{51}
  d_{z,N} &=& \frac{\Gamma(z+1+\frac{3}{2}N)}{\Gamma(z+1+\frac{3}{2}(N-1))} \nonumber \\
  &\times& \left(z+1+\frac{3}{2}N
  \right)^{-3/2}, \qquad z>0, \\ \label{52}
  d_{z,N} &=& \frac{\Gamma(-z-\frac{3}{2}(N-1))}{\Gamma(-z-\frac{3}{2}N)} \nonumber \\
  &\times& \left(-z-1-\frac{3}{2}N
  \right)^{-3/2}, \;\;  z<-\frac{3}{2}N.
\end{eqnarray}
In the Gibbs limit we obtain the usual Maxwell-Boltzmann one-particle distribution function
\begin{equation}\label{64}
 f(\vec{p})|_{z\to\pm\infty} = f_{G}(\vec{p}) = \frac{1}{(2\pi mkT)^{3/2}} \ e^{- \frac{\vec{p}^{2}}{2mkT}},
\end{equation}
where $d_{z,N}|_{z\to\pm\infty}=1$.

Let us investigate the zeroth law of thermodynamics, the principle of additivity and the principle of statistical independence on the base of the exact analytical results for the ideal gas. Divide the system into two parts under the conditions (\ref{25})--(\ref{26a}). Then from Eqs.~(\ref{30})--(\ref{35}) and (\ref{37}), (\ref{39}), it follows that the R\'enyi entropy (\ref{37}) and the Gibbs one (\ref{39}) are nonadditive in finite systems
\begin{eqnarray}\label{54} \nonumber
 S&\neq& S_{1}+S_{2}=S_{G,1}+S_{G,2}+ \\
 &+& k\left[\ln(a_{z_{1},N_{1}}a_{z_{2},N_{2}})-\frac{3}{2}N\right], \\ \label{66}
    S_{G} &\neq&  S_{G,1} +S_{G,2}=k\ln(Z_{G,1}Z_{G,2})+\frac{3}{2}N,
\end{eqnarray}
where
\begin{eqnarray}\label{65}
    Z_{G} &\neq&  Z_{G,1}Z_{G,2} = \frac{V_{1}^{N_{1}}V_{2}^{N_{2}}}{V^{N}}\frac{N!}{N_{1}!N_{2}!}
   \ Z_{G}, \\ \label{65a}
   a_{z,N} &\neq&  a_{z_{1},N_{1}} a_{z_{2},N_{2}}.
\end{eqnarray}
Comparing Eqs.~(\ref{54}), (\ref{66}) and (\ref{10b}), (\ref{11b}), we see that the entropies $S$ and $S_{G}$ are not homogeneous functions. Therefore the zeroth law of thermodynamics is violated both for the R\'enyi and Gibbs statistics of the finite ideal gas in the canonical ensemble. Moreover, the canonical phase space distribution functions for the R\'enyi and Gibbs statistics do not factorize
\begin{eqnarray}\label{55}
\varrho &\neq& \varrho_{1}\varrho_{2} = \frac{1}{Z_{R,1}Z_{R,2}} \\ \nonumber
&\times& \left[1-\frac{1}{kT_{1}(z_{1}+1+\frac{3}{2}N_{1})}
\sum\limits_{i=1}^{N_{1}} \frac{\vec{p}_{i}^{2}}{2m}
\right]^{z_{1}} \times \\ \nonumber &\times&
\left[1-\frac{1}{kT_{2}(z_{2}+1+\frac{3}{2}N_{2})}
\sum\limits_{j=1}^{N_{2}} \frac{\vec{p}_{j}^{2}}{2m}
\right]^{z_{2}}, \\ \label{55a}
\varrho_{G}  &\neq& \varrho_{G,1} \varrho_{G,2} = \frac{1}{Z_{G,1}Z_{G,2}} \ e^{-\frac{1}{kT}\sum\limits_{i=1}^{N} \frac{\vec{p}_{i}^{2}}{2m}},
\end{eqnarray}
where $Z_{R} \neq Z_{R,1}Z_{R,2}$.
Thus the principle of statistical independence is violated, it is polluted by finite-size effects.
This completes the proof of the inequalities (\ref{27}).

Let us now investigate the Euler theorem for the ideal gas in the canonical ensemble. For the Gibbs statistics combining Eqs.~(\ref{32}), (\ref{38}), (\ref{39}), and (\ref{43}), we get
\begin{eqnarray}\label{67}
  TS_{G} + O_{G} &=& \langle H\rangle_{G} +p_{G}V -\mu_{G} N, \\ \label{67a}
  O_{G} &=& kT[\ln N! +N -N\psi(N+1)].
\end{eqnarray}
For the R\'enyi statistics from Eqs.~(\ref{67}), (\ref{67a}) and (\ref{37})--(\ref{46}), we have in the case of $z$ extensive
\begin{eqnarray}\label{53}
  TS + O_{R}^{(e)} &=&  \langle H\rangle +pV +Xz -\mu N, \\ \label{53a} \nonumber
  O_{R}^{(e)} &=& O_{G} + \\
  &+& kT[-\ln a_{z,N} +X_{a}z +A_{\mu}N]
\end{eqnarray}
and in the case of $z$ intensive
\begin{eqnarray}\label{53c}
  TS + O_{R}^{(i)} &=&  \langle H\rangle +pV -\mu N, \\ \label{53d}
  O_{R}^{(i)} &=& O_{G} + kT[-\ln a_{z,N} +A_{\mu}N].
\end{eqnarray}
Thus the Euler theorem is not valid for the finite ideal gas in the canonical ensemble both for the R\'enyi statistics and the Gibbs one. This completes the proof of the inequalities (\ref{19}), (\ref{19bb}).

\subsection{Canonical ideal gas in thermodynamical limit with $z$ being intensive}
Consider the ideal gas in the canonical ensemble for the R\'enyi statistics in the thermodynamic limit with $z$ intensive:
$V\to\infty$, $N\to\infty$ and $v=V/N=const$, $z=const$.
For the sake of convenience let us consider only the case  $z>0$.
Then the phase space distribution function (\ref{30}) and the norm functions (\ref{31}), (\ref{35}) take the form
\begin{equation}\label{68}
\varrho = \frac{1}{Z_{R}} \left[1-\frac{1}{kT(z+1+\frac{3}{2}N)} \sum\limits_{i=1}^{N} \frac{\vec{p}_{i}^{2}}{2m} \right]^{z},
\end{equation}
where
\begin{eqnarray}\label{69}
 Z_{R} &=& \frac{\Gamma(z+1) \ e^{z+1+\frac{3}{2}N}}
 {\sqrt{2\pi} ( z+1+\frac{3}{2} N)^{z+\frac{1}{2}} } \ Z_{G},  \\ \label{70}
 Z_{G}  &=& (gve)^{N} \left(\frac{mkT}{2\pi\hbar^{2}}\right)^{\frac{3}{2}N} = \tilde{Z}_{G}^{N}
\end{eqnarray}
and
\begin{equation}\label{70a}
  a_{z,N} = \frac{\Gamma(z+1) \ e^{z+1+\frac{3}{2}N}}
 {\sqrt{2\pi} (z+1)^{z} ( z+1+\frac{3}{2} N)^{\frac{1}{2}} }.
\end{equation}
In the Gibbs limit $z\to\infty$ and $N=const$ for the distribution function (\ref{68}) and the norm function (\ref{69}), we obtain $\varrho|_{z\to\infty} = \varrho_{G}$ and $Z_{R}|_{z\to\infty} = Z_{G}$, respectively. Then for R\'enyi statistics with $z$ intensive the Gibbs limit commutes with the thermodynamic limit contrary to the Tsallis statistics~\cite{Abe2}.

In the thermodynamic limit with $z$-intensive the mean energy of the system (\ref{32}) and the heat capacity (\ref{47}) are unchanged, $ \langle H\rangle =(3/2)kTN$ and $C_{VzN} =(3/2)kN$, respectively.
However the free energy (\ref{36}) is modified
\begin{eqnarray}\label{71}
  F &=& F_{G}+O(\ln N),   \\ \label{71a}
   F_{G} &=& -kTN \ln \tilde{Z}_{G}, \\ \label{71b}
 O(\ln N) &=& \frac{1}{2} kT \ln\left(z+1+\frac{3}{2}N\right) - \nonumber \\
&-&  kT \ln\left(\frac{\Gamma(z+1) \ e^{z+1}}{\sqrt{2\pi} (z+1)^{z}} \right)  +O(\frac{1}{N}). \;\;\;\;\;\;\;
\end{eqnarray}

The entropy (\ref{37}) in the thermodynamical limit with $z$-intensive can be written as
\begin{eqnarray}\label{72}
  S &=& S_{G}+O(\ln N), \\ \label{72a}
  S_{G} &=& kN  \left( \ln \tilde{Z}_{G}+ \frac{3}{2}\right), \\ \label{72b}
  O(\ln N) &=& -\frac{1}{2} k \ln\left(z+1+\frac{3}{2}N\right) + \nonumber \\
&+&  k \ln\left(\frac{\Gamma(z+1) \ e^{z+1}}{\sqrt{2\pi} (z+1)^{z}} \right) + O(\frac{1}{N}). \;\;\;\;\;
\end{eqnarray}

The pressure (\ref{38}) in the thermodynamical limit with $z$-intensive takes the form
\begin{eqnarray}\label{73}
  p &=& p_{G} = \frac{kT}{v}.
\end{eqnarray}

The chemical potential (\ref{40}) in the thermodynamical limit with $z$-intensive is
\begin{eqnarray}\label{74}
  \mu &=& \mu_{G} + O(\frac{1}{N}), \\ \label{74a}
  \mu_{G} &=& -kT \left( \ln \tilde{Z}_{G}-1\right), \\ \label{74b}
  O(\frac{1}{N}) &=&  \frac{kT}{2N} \frac{1}{1+\left(\frac{3}{2}\frac{N}{z+1}\right)^{-1}} + O(\frac{1}{N^{2}}).
\end{eqnarray}

The variable (\ref{44}) in the thermodynamical limit with $z$-intensive can be written as
\begin{eqnarray}\label{75}
  X &=& X_{0}N + O(N^{0}), \qquad X_{0}=0, \\ \label{75a}
 O(N^{0}) &=& kT [ \psi(z+1)-\ln(z+1) ] + \nonumber \\
&+&  \frac{kT}{z+1}\left[1-\frac{1}{2}\left(\frac{3}{2}\frac{N}{z+1}\right)^{-1}\right] + O(\frac{1}{N^{2}}). \;\;\;\;\;\;\;\;\;
\end{eqnarray}

The functions (\ref{71})--(\ref{75}) were obtained by expanding them in power series for small $1/N$.
For extensive functions the term proportional to $N$ and for intensive functions the term proportional to $N^{0}$ were kept. Note that the chemical potential (\ref{74}) and the thermodynamic variable (\ref{75}) are not consistent with the thermodynamical relations (\ref{22}) with respect to the thermodynamical potential (\ref{71}). In the Gibbs limit for the thermodynamic variable (\ref{75}), we have $X|_{z\to\infty}=0$.

In the thermodynamical limit with $z$-intensive the $N$-particle distribution function (\ref{48}) of the classical ideal gas for R\'{e}nyi statistics in the canonical ensemble is unchanged
\begin{eqnarray}\label{76}
    f(\vec{p}_{1},\ldots,\vec{p}_{N}) &=&
    \frac{(gve)^{N}}{h^{3N}Z_{R}}  \\ &\times& \left[1-
    \frac{1}{kT(z+1+\frac{3}{2}N)}
    \sum\limits_{i=1}^{N} \frac{\vec{p}_{i}^{2}}{2m} \right]^{z}.
\nonumber
\end{eqnarray}
However the single-particle distribution function (\ref{50}) is reduced to the usual Maxwell-Boltzmann form
\begin{equation}\label{77}
  f(\vec{p}) =  f_{G}(\vec{p}) = \frac{1}{(2\pi mkT)^{3/2}} \
   e^{-\frac{\vec{p}^{2}}{2mkT}}.
\end{equation}
In the Gibbs limit $z\to\infty$ and $N=const$ for the $N$-particle distribution function (\ref{76}), we have $ f(\vec{p}_{1},\ldots,\vec{p}_{N})|_{z\to\infty}= f_{G}(\vec{p}_{1},\ldots,\vec{p}_{N})$ (\ref{63}).

Let us also investigate the zeroth law of thermodynamics, the principle of additivity and the principle of statistical independence for the ideal gas in the thermodynamical limit with $z$-intensive. Divide the system into two parts under the conditions (\ref{25}), (\ref{26a}). Then from Eqs.~(\ref{72})--(\ref{72a}) and (\ref{70}), it follows that the R\'enyi entropy (\ref{72}) and the Gibbs one (\ref{72a}) are additive (extensive)
\begin{eqnarray}\label{78}
S &=& S_{1}+S_{2}=S_{G,1}+S_{G,2}, \\ \label{78a}
 S_{G} &=& S_{G,1}+S_{G,2},
\end{eqnarray}
where $\tilde{Z}_{G}=\tilde{Z}_{G,1}=\tilde{Z}_{G,2}$.
Here terms proportional to $N$ were kept only. Therefore for the R\'enyi and Gibbs statistics in the thermodynamical limit with $z$-intensive the zeroth law of thermodynamics is satisfied. Comparing Eqs.~(\ref{78}), (\ref{78a}) and (\ref{10b}), (\ref{11b}), we see that the entropies $S$ and $S_{G}$ are homogeneous functions. The phase space distribution function for the Gibbs statistics (\ref{58}), (\ref{55a}) factorizes, however, the phase space distribution function for the R\'enyi statistics (\ref{55}) does not factorize
\begin{eqnarray}\label{79}
\varrho &\neq& \varrho_{1}\varrho_{2} = \frac{1}{Z_{R,1}Z_{R,2}} \\ \nonumber
&\times& \left[1-\frac{1}{kT(z+1+\frac{3}{2}N_{1})}
\sum\limits_{i=1}^{N_{1}} \frac{\vec{p}_{i}^{2}}{2m}
\right]^{z} \times \\ \nonumber &\times&
\left[1-\frac{1}{kT(z+1+\frac{3}{2}N_{2})}
\sum\limits_{j=1}^{N_{2}} \frac{\vec{p}_{j}^{2}}{2m}
\right]^{z}, \\ \label{79a}
\varrho_{G}  &=& \varrho_{G,1} \varrho_{G,2} = \frac{1}{Z_{G}} \ e^{-\frac{1}{kT}\sum\limits_{i=1}^{N} \frac{\vec{p}_{i}^{2}}{2m}},
\end{eqnarray}
where $Z_{R} \neq Z_{R,1}Z_{R,2}$ and $Z_{G} = Z_{G,1}Z_{G,2}$. This means that the principle of statistical independence is not satisfied for the R\'enyi statistics at finite $N_1$ and $N_2$.

Combining Eqs.~(\ref{67}), (\ref{67a}), (\ref{53c}), and (\ref{53d}), we obtain the Euler theorem
\begin{eqnarray}\label{81}
  TS &=& \langle H\rangle + pV - \mu N, \quad O_{R}^{(i)}=O(\ln N), \;\;\;\;  \\ \label{81a}
  TS_{G} &=& \langle H\rangle_{G} +p_{G}V -\mu_{G} N, \quad O_{G}=0.
\end{eqnarray}
Here only terms proportional to $N$ are kept.
Thus for the R\'{e}nyi and Gibbs statistics in the thermodynamical limit with $z$-intensive the zeroth law of thermodynamics, the principle of additivity and the Euler theorem are satisfied. They are valid whenever the entropy (\ref{11b}) is a homogeneous function in the thermodynamic limit.

\subsection{Ideal gas in thermodynamical limit with $z$ being extensive}

Consider now the ideal gas in the canonical ensemble for the R\'enyi statistics in the thermodynamic
limit with $z$ extensive: $V\to\infty$, $N\to\infty$, $z\to\pm\infty$ and $v=V/N=const$,
$\tilde{z}=z/N=const$. Then the phase space distribution function (\ref{30}) and the norm
functions (\ref{31}), (\ref{33}), and (\ref{34}) take the form
\begin{equation}\label{82}
\varrho = \frac{1}{Z_{R}} \left[1-\frac{1}{kT(\tilde{z}+\frac{3}{2})N} \sum\limits_{i=1}^{N}
\frac{\vec{p}_{i}^{2}}{2m} \right]^{\tilde{z}N},
\end{equation}
where
\begin{eqnarray}\label{83}
 Z_{R} &=&  e^{\frac{3}{2}N}\left(1+\frac{3}{2\tilde{z}}\right)^{-\tilde{z}N-1/2}  Z_{G}, \\ \label{83a}
 a_{z,N} &=&  e^{\frac{3}{2}N}\left(1+\frac{3}{2\tilde{z}}\right)^{-1/2}.
\end{eqnarray}
The partition function $Z_{G}$ is given in Eq.~(\ref{70}). In the Gibbs limit $\tilde{z}\to\pm\infty$ and $N=const$ for the distribution function (\ref{82}) and the norm function (\ref{83}), we obtain $\varrho|_{\tilde{z}\to\pm\infty} = \varrho_{G}$ and $Z_{R}|_{\tilde{z}\to\pm\infty} = Z_{G}$, respectively. Then in R\'enyi statistics with $z$ extensive the Gibbs limit is commutative with the thermodynamic limit contrary to the Tsallis statistics~\cite{Abe2}.

The mean energy of the system (\ref{32}) and the heat capacity (\ref{47}) are unchanged, $ \langle H\rangle =(3/2)kTN$ and $C_{VzN} =(3/2)kN$, respectively. The free energy (\ref{36}) in the thermodynamical limit with $z$ extensive can be written as
\begin{eqnarray}\label{84}
  F &=& F_{G} + O(N^{0}), \\ \label{84a}
 O(N^{0}) &=& \frac{kT}{2} \ln\left(1+\frac{3}{2\tilde{z}}\right) + O(\frac{1}{N}).
\end{eqnarray}
The entropy (\ref{37}) and the pressure (\ref{38}) in the thermodynamical limit with $z$-extensive are
\begin{eqnarray}\label{85}
  S &=& S_{G} + O(N^{0}), \\ \label{85a}
 O(N^{0}) &=& -\frac{k}{2} \ln\left(1+\frac{3}{2\tilde{z}}\right) + O(\frac{1}{N}), \\ \label{86}
  p &=& p_{G} = \frac{kT}{v}.
\end{eqnarray}

The chemical potential (\ref{40}) and the variable (\ref{44}) in the thermodynamical limit with $z$-extensive are
\begin{eqnarray}\label{87}
  \mu &=& \mu_{G} +  O(\frac{1}{N}), \\ \label{87a}
   O(\frac{1}{N}) &=& \frac{kT}{2N} \frac{1}{1+\left(\frac{3}{2\tilde{z}}\right)^{-1}} + O(\frac{1}{N^{2}})
 \end{eqnarray}
and
 \begin{eqnarray}\label{88}
  X &=& X_{0}N^{0}+ O(\frac{1}{N}), \qquad X_{0}=0, \\ \label{88a}
   O(\frac{1}{N}) &=& \frac{kT}{2\tilde{z}N} \frac{1}{1+\left(\frac{3}{2\tilde{z}}\right)^{-1}} + O(\frac{1}{N^{2}}).
\end{eqnarray}
Here for the extensive functions (\ref{84}), (\ref{85}) only terms proportional to $N$ are kept,
and for the intensive functions (\ref{87}), (\ref{88}) terms proportional to $N^{0}$.
Note that the chemical potential (\ref{87}) obtained in the thermodynamic limit does not agree with the thermodynamical relation (\ref{22}) and the thermodynamical potential (\ref{84}).

The $N$-particle distribution function (\ref{48}) of the classical ideal gas for R\'{e}nyi statistics in the thermodynamical limit with $z$ extensive takes the form
\begin{eqnarray}\label{89}
    f(\vec{p}_{1},\ldots,\vec{p}_{N}) &=& \frac{(gve)^{N}}{h^{3N}Z_{R}}  \\ &\times& \left[1-
    \frac{1}{kT(\tilde{z}+\frac{3}{2})N} \sum\limits_{i=1}^{N} \frac{\vec{p}_{i}^{2}}{2m} \right]^{\tilde{z}N}.
\nonumber
\end{eqnarray}
At the same time the single-particle distribution function (\ref{50}) is reduced to the Maxwell-Boltzmann form
\begin{equation}\label{90}
  f(\vec{p}) =  f_{G}(\vec{p}) =  \frac{1}{(2\pi mkT)^{3/2}} \ e^{-\frac{\vec{p}^{2}}{2mkT}},
\end{equation}
where $d_{z,N}=1$. In the Gibbs limit $\tilde{z}\to\pm\infty$ and $N=const$ for the $N$-particle distribution function (\ref{89}), we have $ f(\vec{p}_{1},\ldots,\vec{p}_{N})|_{\tilde{z}\to\pm\infty}= f_{G}(\vec{p}_{1},\ldots,\vec{p}_{N})$ (\ref{63}).

Finally, let us investigate the zeroth law of thermodynamics, the principle of additivity and the principle of statistical independence for the ideal gas in the thermodynamical limit with $z$ extensive. Divide the system into two parts under the conditions (\ref{25}), (\ref{26}). Then from Eqs.~(\ref{82}), (\ref{83}) and (\ref{85}), (\ref{85a}), it follows that the R\'enyi entropy (\ref{85}) and the Gibbs one (\ref{72a}) are additive but the distribution function (\ref{82}) does not factorize
\begin{equation}\label{91}
S = S_{1}+S_{2}=S_{G,1}+S_{G,2}
\end{equation}
and
\begin{eqnarray}\label{91a} \nonumber
\varrho &\neq& \varrho_{1}\varrho_{2} =  \\ \nonumber
&=& \frac{1}{Z_{R}} \left[1-\frac{1}{kT(\tilde{z}+\frac{3}{2})N_{1}} \sum\limits_{i=1}^{N_{1}} \frac{\vec{p}_{i}^{2}}{2m} \right]^{\tilde{z}N_{1}} \times \\  &\times& \left[1-\frac{1}{kT(\tilde{z}+\frac{3}{2})N_{2}} \sum\limits_{j=1}^{N_{2}} \frac{\vec{p}_{j}^{2}}{2m} \right]^{\tilde{z}N_{2}},
\end{eqnarray}
where the norm function (\ref{83}) factorizes, $Z_{R}=Z_{R,1}Z_{R,2}$.
The phase space distribution functioni, however, does not factorize.
Therefore statistical independence is not achieved.
Combining Eqs.~(\ref{85})--(\ref{88}), we obtain the Euler theorem for $z$ extensive
\begin{equation}\label{92}
  TS = \langle H\rangle +pV +Xz -\mu N,
\end{equation}
where $X=0$. In Eqs.~(\ref{91}), (\ref{92}) was kept only the terms proportional to the first power of $N$. Thus for the R\'{e}nyi statistics in the thermodynamical limit with $z$ extensive the zeroth law of thermodynamics, the principle of additivity and the Euler theorem are satisfied. They are valid whenever the entropy
(\ref{85}) is equivalent with (\ref{10b}) and it is a homogeneous function of the first degree.

By the exact analytical results for the ideal gas, it was proved that the R\'enyi statistics in the canonical ensemble is thermodynamically self-consistent in the thermodynamical limit, for $z$ being either extensive or intensive.
In this case the ideal gas in the canonical ensemble for R\'enyi statistics is equivalent with
the ideal gas for the Gibbs statistics, except for the N-particle phase space distribution function,
which on the other hand lacks the physical sense of a thermodynamical limit.

%%%%%%%%%%%%%% SECTION 3 %%%%%%%%%%%%%%%%%%%
\section{Microcanonical ensemble $(E,V,z,N)$}
\subsection{General formalism}
Consider the equilibrium statistical ensemble of the closed energetically isolated systems of $N$ particles at the constant volume $V$ and the constant thermodynamic coordinate $z$.
Suppose, these systems have identical energy $E$ within $\Delta E\ll E$; then the phase space distribution function $\varrho$ be distinct from zero only in that region of phase space, $D$,
which is defined by the inequality $E\leq H\leq E +\Delta E$,
with $H$ being the Hamiltonian of the system.

To express the equilibrium phase space distribution function through the macroscopic variables of state of the microcanonical ensemble we use the thermodynamical method based on the fundamental equation of thermodynamics~\cite{Parv1}. The fundamental equation of thermodynamics at constant values of $E,V,z,N$ can be written as
\begin{equation}\label{93}
     (dS)_{E,V,z,N} = 0.
\end{equation}
Using Eq.~(\ref{1}), we obtain
\begin{equation}\label{94}
 dS = \int\limits_{D}  d\Gamma \frac{\delta S}{\delta \varrho}
    d\varrho + \frac{\partial S}{\partial z}dz = 0.
\end{equation}
Since in the microcanonical ensemble these differentials vanish, $dz=d\varrho=0$, we obtain
\begin{equation}\label{95}
    \frac{\delta S}{\delta \varrho} = \alpha_{1}, \qquad
     \frac{\partial S}{\partial z} = \alpha_{2},
\end{equation}
where $\alpha_{1}$ and $\alpha_{2}$ are certain constants. Substituting Eq.~(\ref{1}) for the R\'enyi entropy in the first equation of (\ref{95}) and in its expectation value, we get
\begin{eqnarray}\label{96}
  \frac{\delta S}{\delta \varrho} &=&  -k (z+1) \varrho^{1/z} e^{\frac{S}{kz}} =
  \alpha_{1}, \\ \label{97}
 \int\limits_{D}  d\Gamma \varrho \frac{\delta S}{\delta \varrho} &=& -k (z+1)= \alpha_{1}.
\end{eqnarray}
Combining Eqs.~(\ref{96}), (\ref{97}), and (\ref{2}), we obtain the phase space distribution function and the statistical weight for the R\'enyi statistics in the framework of the classical statistical mechanics~\cite{ParvBiro}
\begin{equation}\label{98}
\varrho = \frac{1}{W} = \varrho_{G},
\end{equation}
where
\begin{equation}\label{99}
W = \int\limits_{D} d\Gamma = \int d\Gamma \Delta(H-E).
\end{equation}
Then the R\'{e}nyi entropy (\ref{1}) is reduced to the familiar expression
\begin{equation}\label{100}
    S=k\ln W \equiv S_{G},
\end{equation}
where $S_{G}$ is the Gibbs entropy. Since the R\'enyi entropy (\ref{100}) is a thermodynamical potential of the microcanonical ensemble and it is equivalent with the Gibbs entropy, and it is independent of the variable of the state $z$, we obtain that in the microcanonical ensemble the R\'{e}nyi statistics is equivalent with the usual Boltzmann-Gibbs statistics and it is independent of the variable of the state $z$ ($\alpha_2=0$).
Therefore in the microcanonical ensemble we have only three variables of state $(E,V,N)$. From Eq.~(\ref{98}) it follows that the expectation value of a dynamical variable $A$ (\ref{3}) can be written as
\begin{equation}\label{101}
 \langle A\rangle = \int\limits_{D} d\Gamma \varrho A=
  \frac{1}{W}\int d\Gamma \Delta(H-E) A ,
\end{equation}
where $\langle A\rangle=\langle A\rangle_{G}$. For the distribution function (\ref{98}) and the entropy (\ref{100}), we have the fundamental equation of thermodynamics
\begin{equation}\label{102}
    TdS= dE + p dV - \mu dN,
\end{equation}
where
\begin{eqnarray}\label{103}
  \frac{1}{T} &=& \left(\frac{\partial S}{\partial E}\right)_{V,N}=\frac{1}{T_{G}}, \\ \label{104}
  p &=& T\left(\frac{\partial S}{\partial V}\right)_{E,N}= p_{G}, \\ \label{105}
  \mu &=& -T\left(\frac{\partial S}{\partial N}\right)_{E,V}=\mu_{G}.
\end{eqnarray}
Here $\langle H\rangle=E$ and $X=T(\partial S/\partial z)=0$.

The fundamental equation of thermodynamics (\ref{102}) provides the first and second laws of thermodynamics
\begin{equation}\label{106}
    \delta Q=TdS, \;\; \delta Q= d\langle H\rangle + pdV  -\mu dN.
\end{equation}
The heat capacity, $\delta Q =C dT$, in the microcanonical ensemble at the constant values of $E,V,z,N$ can be written as
\begin{eqnarray}\label{107}
    C_{EVN} &=& \frac{1}{\left(\frac{\partial T}{\partial E}\right)_{V,N}}=
    -\frac{1}{T^{2}\left(\frac{\partial^{2} S}{\partial
    E^{2}}\right)_{V,N}},
\end{eqnarray}
where $ C_{EVN}= C_{EVN}^{G}$.

Let us prove the zeroth law of thermodynamics for the R\'enyi statistics in the microcanonical ensemble. Divide the system into two parts, $H = H_{1}+H_{2}$, under the conditions
\begin{equation}\label{1x}
  E = E_{1}=E_{2}, \;  N = N_{1}+N_{2}, \; V=V_{1}+V_{2}.
\end{equation}
The heat transfer $\delta Q$ is additive. Therefore, from the second law of thermodynamics (\ref{106}) it follows that the zeroth law, $T = T_{1}=T_{2}$, is valid whenever the total differential of entropy is additive $dS=dS_{1}+dS_{2}$ (see Eqs.~(\ref{1b}), (\ref{2b})). Let us show that Eq.~(\ref{2b}) is valid only if the entropy $S$ is additive function
\begin{equation}\label{2x}
    S(E,V,N) = S_{1}(E_{1},V_{1},N_{1}) + S_{2}(E_{2},V_{2},N_{2}).
\end{equation}
Differentiating Eq.~(\ref{2x}) with respect to variables of state $E,V$ and $N$, we obtain
\begin{eqnarray}\label{3x}
  \frac{\partial S}{\partial E} &=& \frac{\partial S_{1}}{\partial E_{1}} =   \frac{\partial S_{2}}{\partial E_{2}},
  \qquad E = E_{1}+E_{2},\\ \label{4x}
   \frac{\partial S}{\partial V} &=& \frac{\partial S_{1}}{\partial V_{1}} =   \frac{\partial S_{2}}{\partial V_{2}},
   \qquad V=V_{1}+V_{2},\\ \label{5x}
   \frac{\partial S}{\partial N} &=& \frac{\partial S_{1}}{\partial N_{1}} =   \frac{\partial S_{2}}{\partial N_{2}}, \qquad N = N_{1}+N_{2}
\end{eqnarray}
or
\begin{eqnarray}\label{6x}
  T &=& T_{1} = T_{2}, \qquad E = E_{1}+E_{2},\\ \label{7x}
   p &=& p_{1} =  p_{2}, \qquad V=V_{1}+V_{2},\\ \label{8x}
    \mu &=& \mu_{1} = \mu_{2}, \qquad N = N_{1}+N_{2}.
\end{eqnarray}
Then from Eqs.~(\ref{3x})--(\ref{5x}) and the differentials of Eq.~(\ref{1x}), we get
\begin{equation}\label{9x}
    dS = dS_{1}+dS_{2}.
\end{equation}
Therefore the zeroth law of thermodynamics, $T = T_{1}=T_{2}$, in the microcanonical ensemble for the R\'enyi statistics  is satisfied whenever the entropy is additive, $S=S_{1}+S_{2}$.

Let us show the correspondence in the microcanonical ensemble between the additivity of entropy (\ref{2x}) and its homogeneity. Consider the R\'{e}nyi (Gibbs) entropy in the microcanonical ensemble be a first order homogeneous function of variables $E,V,N$
\begin{eqnarray}\label{10x}
  S(E,\lambda V,\lambda N) &=& \lambda S(E,V,N), \quad \lambda=\frac{1}{N}.
\end{eqnarray}
Divide the system into two parts under the conditions (\ref{1x}). For $\lambda$, we have Eq.~(\ref{12b}).
Then the intensive specific variables correspond to the additive (extensive) variables
\begin{eqnarray}\label{11x}
\lambda E=\lambda_{1} E_{1}=\lambda_{2} E_{2}, \qquad \lambda V=\lambda_{1} V_{1}=\lambda_{2} V_{2},
\end{eqnarray}
where $\lambda_{1}=1/N_{1}$ and $\lambda_{2}=1/N_{2}$. Under the conditions of Eqs.~(\ref{2x}), (\ref{10x}), and (\ref{11x}), it follows that the specific entropy $\lambda S$ is intensive function
\begin{eqnarray}\label{12x} \nonumber
   \lambda S(E,V,N) &=& \lambda_{1} S_{1}(E_{1},V_{1},N_{1}) = \\
   &=& \lambda_{2} S_{2}(E_{2},V_{2},N_{2}).
\end{eqnarray}
Conversely, from Eqs.~(\ref{12x}), (\ref{10x}), and (\ref{11x}), we obtain the additivity of entropy (\ref{2x}). Thus the zeroth law of thermodynamics for R\'enyi statistics in the microcanonical ensemble is valid whenever the entropy (or the thermodynamic potential) is a homogeneous function of first degree of the extensive variables of state~\cite{Parv1,Parv2}.

If the entropy is a homogeneous function of degree $1$ (\ref{10x}), then $S$ satisfies the Euler theorem and the Gibbs-Duhem relation. Differentiating Eqs.~(\ref{10x}) with respect to $\lambda$, and putting $\lambda=1$, we obtain the well-known Euler theorem for the homogeneous functions:
\begin{equation} \label{112}
  E\left(\frac{\partial S} {\partial E} \right)_{V,N} +
    V\left(\frac{\partial S} {\partial V} \right)_{E,N} +
    N\left(\frac{\partial S} {\partial N} \right)_{E,V} = S.
\end{equation}
Using the thermodynamic relations (\ref{103})--(\ref{105}) for the isolated thermodynamic system $(E,V,N)$, we get the Euler theorem~\cite{Prigogine}:
\begin{equation} \label{113}
 T S=E+p V -\mu N.
\end{equation}
Applying the differential operator with respect to the variables of state $(E,V,N)$ on Eq.~(\ref{113}), we obtain the fundamental equation of thermodynamics (\ref{102}) and the Gibbs-Duhem relation~\cite{Prigogine}
\begin{equation}\label{114}
    S dT = V dp -N d\mu.
\end{equation}
Equation (\ref{114}) means that the variables $T$, $\mu$ and $p$ are not independent. The fundamental equation of
thermodynamics (\ref{102}) provides the first and the second laws of thermodynamics (\ref{106}). Therefore in the microcanonical ensemble the Euler theorem (\ref{113}) and the Gibbs-Duhem relation (\ref{114}) result from the homogeneity property of the entropy (\ref{10x}).

Dividing the finite system of the microcanonical ensemble into two dynamically independent subsystems, $H=H_{1}+H_{2}$, under the conditions (\ref{1x}), we have a convolution of the statistical weight $W$
\begin{eqnarray}\label{108}
     &&W(E,V,N) = \frac{N_{1}!N_{2}!}{N!} \frac{V_{1}^{N_{1}}V_{2}^{N_{2}}}{V^{N}} \times
    \qquad \;\;\;\;\;\;\;\;\;\;\;\;\; \nonumber \\
     &\times& \int\limits_{0}^{E} dE_{1} W_{1}(E_{1},V_{1},N_{1}) W_{2}(E-E_{1},V_{2},N_{2}), \;\;\;\;\;
\end{eqnarray}
where $W_{i} = \int\Delta(H_{i}-E_{i}) d\Gamma_{i}$. Hence the statistical weight (\ref{99}) and the distribution function (\ref{98}) do not factorize, $W\neq W_{1}W_{2}$ and $\varrho \neq \varrho_{1}\varrho_{2}$, respectively. Therefore the  R\'{e}nyi (Gibbs) entropy (\ref{100}) is nonadditive (nonextensive) function and the proper temperature (\ref{103}) is not an intensive variable
\begin{equation}\label{109}
  S \neq S_{1}+S_{2}, \qquad  T\neq T_{1} \neq T_{2}.
\end{equation}
Thus the Gibbs and R\'{e}nyi statistics for finite systems violate the zeroth law of thermodynamics (see Eq.~(\ref{109})). Moreover, the Euler theorem is also not valid
\begin{equation}\label{110}
  TS \neq \langle H\rangle +pV -\mu N.
\end{equation}
For finite values of the variables of state $E,V,N$ the thermodynamical potential (\ref{100}) of the microcanonical ensemble is not a homogenous function of first order. The proof of inequalities (\ref{109}), (\ref{110}) is given below on the base of the ideal gas.

In the general case, the R\'{e}nyi (Gibbs) entropy (\ref{100}) of the variables of state is not a homogeneous function. It can become homogeneous only in the thermodynamic limit, $N\rightarrow\infty$, $E\rightarrow\infty$, $V\rightarrow\infty$ and $v=V/N =\mathrm{const}$, $\tilde{z}=z/N =\mathrm{const}$. This means that the functions of variables of state can be expand in power series with the small parameter $1/N$ $(N\gg 1)$ at large finite values of the variables $E,V$. Then the extensive variables $\mathcal{A}$ and the intensive ones $\phi$ can be written $(\alpha> 0)$ as
\begin{eqnarray} \label{13x} \nonumber
   \mathcal{A}(E,V,N) &=& N[a(\varepsilon,v) + O(N^{-\alpha})] \\
    &\stackrel{as} =&  N a(\varepsilon,v), \\
     \label{14x} \nonumber
\phi(E,V,N) &=& \phi(\varepsilon,v) + O(N^{-\alpha}) \\
      &\stackrel{as} =& \phi(\varepsilon,v).
\end{eqnarray}
where $\varepsilon=E/N$ is the specific energy, $v=V/N$ is the specific volume, and $a =\mathcal{A} /N$ is the specific $\mathcal{A}$. Comparing (\ref{13x}) and (\ref{10x}), it follows that the entropy is a homogeneous function of degree $1$ only in the thermodynamic limit. Thus the zeroth law of thermodynamics, the Euler theorem and the Gibbs-Duhem relation for the R\'enyi (Gibbs) statistics can be satisfied only in the thermodynamic limit.

\subsection{Ideal gas in microcanonical ensemble}
The thermodynamic properties of the R\'{e}nyi (Gibbs) statistics in the microcanonical ensemble $(E,V,N)$ can be investigated by the exact solved analytical functions of the variables of state for the classical nonrelativistic ideal gas. The main relations for the ideal gas of the Gibbs statistics can be found in~\cite{Parv1}. The phase space distribution function (\ref{98}) and the statistical weight (\ref{99}) are given by~\cite{Das}
\begin{eqnarray} \label{120}
 \varrho = \frac{1}{W},  && W(E,V,N) =  \frac{V^{N}}{N!h^{3N}}\int d^{3}p_{1}\ldots
 d^{3}p_{N}   \nonumber \\
 \nonumber &\times& \delta\left(\sum\limits_{i=1}^{N}\frac{\vec{p}_{i}^{2}}{2m}-E\right)
 = \\ &=& \frac{V^{N}}{N!}
 \left(\frac{m} {2\pi \hbar ^{2}} \right)^{\frac{3} {2}
N}  \frac{E ^{\frac{3} {2} N-1}} {\Gamma (\frac{3} {2} N)},
\end{eqnarray}
where $m$ is the particle's rest mass. The entropy (\ref{100}) and the temperature (\ref{103}) take the form
\begin{eqnarray} \label{121}
  S &=& S_{G} = k\ln\left[ \frac{V^{N}}{N!}
 \left(\frac{m} {2\pi \hbar ^{2}} \right)^{\frac{3} {2}
N} \frac{E ^{\frac{3} {2} N-1}} {\Gamma (\frac{3} {2} N)}\right], \;\;\;
  \\ \label{122}
  T &=& T_{G} = \frac{E}{(\frac{3}{2}N-1)k}.
\end{eqnarray}
The pressure (\ref{104}), the chemical potential (\ref{104}) and the heat capacity (\ref{107}) can be written as
\begin{eqnarray} \label{123}
  p &=& p_{G} = \frac{N}{V} kT, \\ \label{124}
  \mu &=& \mu_{G} = -kT\ln\left[V\left(\frac{mE}{2\pi\hbar^{2}}\right)^{3/2}\right]+ \nonumber \;\;\; \\
  &+& kT \left[\psi(N+1)+\frac{3}{2}\psi\left(\frac{3}{2}N\right)\right],   \\ \label{125}
  C_{EVN} &=& C_{EVN}^{G} = \left( \frac{3}{2}N-1\right) k = \frac{E}{T}.
\end{eqnarray}

Divide the system into two parts under the conditions (\ref{1x}). Then from Eqs.~(\ref{120})--(\ref{122}), it follows that the R\'enyi (Gibbs) entropy (\ref{121}) is nonadditive, the temperature (\ref{122}) is nonintensive and the phase space distribution function does not factorize
\begin{eqnarray}\label{15x}
  S &\neq& S_{1} + S_{2} = k\ln(W_{1}W_{2}), \quad   T \neq T_{1}\neq T_{2}, \;\;\; \\ \label{16x}
  \varrho &\neq& \varrho_{1} \varrho_{2} = \frac{1}{W_{1}W_{2}},
\end{eqnarray}
where $W \neq W_{1}W_{2}$. Using Eqs.~(\ref{121})--(\ref{124}), we have
\begin{eqnarray} \label{17x}
 TS+O_{R} &=& E+pV-\mu N, \\ \label{18x} \nonumber
  O_{R} &=& E \left(1+\frac{N}{\frac{3}{2}N-1}\right)+ \\ \nonumber
  &+& \frac{E}{\frac{3}{2}N-1}
  \ln\left(EN!\Gamma(\frac{3}{2}N)\right)- \\
 &-& \frac{EN}{\frac{3}{2}N-1} \left[\psi(N+1)+\frac{3}{2}\psi(\frac{3}{2}N)\right]. \;\;\;\;\;\;
\end{eqnarray}
Thus the zeroth law of thermodynamics, the principle of additivity and the Euler theorem are not satisfied in the microcanonical ensemble for the finite ideal gas of the R\'{e}nyi (Gibbs) statistics. This completes the proof of the inequalities (\ref{109}), (\ref{110}).

Consider now the ideal gas in the microcanonical ensemble for the R\'enyi statistics in the thermodynamic limit
$E\to\infty$, $V\to\infty$, $N\to\infty$ and $\varepsilon=E/N=const$, $v=V/N=const$. Then the phase space distribution function and the statistical weight (\ref{120}) can be written as
\begin{equation}\label{128}
    \varrho=\frac{1}{W}, \quad W= w^{N}, \quad w= ve \left( \frac{m\varepsilon e}{3\pi \hbar^{2}}\right)^{3/2}.
\end{equation}
The entropy (\ref{100}), the temperature (\ref{103}), the pressure (\ref{104}), the chemical potential (\ref{104}) and the heat capacity (\ref{107}) in the thermodynamic limit are found to be
\begin{eqnarray}\label{129}
  S &=& S_{G} = kN \ln w , \\ \label{130}
  T &=& T_{G}=\frac{2}{3} \frac{\varepsilon}{k},  \\ \label{131}
  p &=& p_{G}= \frac{kT}{v} = \frac{2}{3} \frac{\varepsilon}{v},\\ \label{132}
  \mu &=& \mu_{G}=-kT  \left(\ln w -\frac{5}{2} \right),  \\ \label{133}
  C_{EVN} &=& C_{EVN}^{G} = \frac{3}{2}kN=N\frac{\varepsilon}{T}.
\end{eqnarray}
In the thermodynamical limit the Euler theorem (\ref{17x}), (\ref{18x}) is valid
\begin{equation}\label{134}
    Ts=\varepsilon +pv -\mu,
\end{equation}
where $s=S/N$.

It is easy to prove that in the thermodynamic limit under the conditions $E=E_{1}+E_{2},V=V_{1}+V_{2}$ and $N=N_{1}+N_{2}$ or $\varepsilon=\varepsilon_{1}=\varepsilon_{2}$, $v=v_{1}=v_{2}$ the R\'{e}nyi entropy (\ref{129}) is additive, the phase space distribution function (\ref{128}) factorizes and the temperature (\ref{130}) is intensive, i.e., the zeroth law of thermodynamics is valid
\begin{equation}\label{119}
 S=S_{1}+S_{2}, \qquad  T=T_{1}=T_{2}.
\end{equation}
Here $\varrho = \varrho_{1} \varrho_{2}$ and $W = W_{1}W_{2}$. Therefore the R\'{e}nyi (Gibbs) statistics in the microcanonical ensemble completely satisfies all requirements of the equilibrium thermodynamics.
Let us note that the zeroth law of thermodynamics, the principle of additivity and the Euler theorem for the R\'{e}nyi (Gibbs) statistics in the microcanonical ensemble are satisfied due to the fact that in the thermodynamic limit the thermodynamical potential of the microcanonical ensemble, the entropy (\ref{129}),
becomes a homogeneous function of first degree of the extensive variables of state.

%%%%%%%%%%%%%%%%%%%%%%%%%% SECTION 5 %%%%%%%%%%%%%%%%%%%%
\section{Conclusions}

To conclude, this paper has explored the R\'enyi statistics in the canonical and  microcanonical ensembles in the general case and for the classical ideal gas.
The exact analytical results for the ideal gas were obtained both for a finite system and in the thermodynamic limit.
In the canonical ensemble the ideal gas in a thermodynamic limit was studied in two cases: when the variable $z$ is extensive and when it is intensive.
The problem of the connection between the statistical mechanics based on the R\'enyi entropy and the equilibrium thermodynamics in the canonical and microcanonical ensembles was analyzed.

Here we summarize the main results of this study. The phase space distribution functions for the canonical and microcanonical ensembles were derived by the method based on the fundamental equation of thermodynamics instead of the Jaynes principle. Both these methods give the same results. For the R\'enyi statistics in the canonical ensemble we obtain the power-law phase space distribution function fixed by two norm functions instead of one partition function given in the Gibbs statistics. However in the microcanonical ensemble we have the usual equiprobability distribution function. The main thermodynamical relations and laws for a statistical ensemble of the R\'enyi statistics were obtained from the ensemble averages and the phase space distribution function. The connection of the R\'enyi statistics and the equilibrium thermodynamics was established.
By the exact analytical results for the ideal gas, it was shown that the zeroth law of thermodynamics, the principle of additivity, the Euler theorem and the Gibbs-Duhem relation are valid only if the thermodynamical potential of the statistical ensemble is a homogeneous function of degree $1$ of its extensive variables of state.
For finite systems the thermodynamical potential is an inhomogeneous function and the
R\'enyi (Gibbs) statistics is neither in the canonical nor in the microcanonical ensemble
is thermodynamically self-consistent.
The thermodynamical potential becomes a homogenous function of first degree only in the thermodynamic limit.
In this  case it is additive, and the R\'enyi (Gibbs) statistics is thermodynamically self-consistent.

It was further shown that the thermodynamical potential of the microcanonical ensemble,
the R\'enyi entropy, is equivalent with the Gibbs entropy and is independent of the
variable of state $z$.
Therefore in the microcanonical ensemble the R\'enyi statistics completely coincides with the usual Boltzmann-Gibbs statistics. It was found that for finite systems in the canonical ensemble the entropic index $z$ must be an extensive variable of state for the Euler theorem being consistent with the
fundamental equation of thermodynamics and the Gibbs-Duhem relation. It was revealed that in the canonical ensemble in the thermodynamical limit both for $z$ extensive and $z$ intensive the ideal gas of the R\'enyi statistics is equivalent with the ideal gas of the Gibbs statistics, except for the phase space distribution function,
which nevertheless in the thermodynamic limit lacks physical sense.
Evidently, the R\'enyi statistics in the canonical and microcanonical ensembles is
thermodynamically self-consistent whenever it coincides with the usual Boltzmann-Gibbs statistics.
%Therefore following the principle of simplicity we conclude that the R\'enyi statistics is unnecessary in the equilibrium statistical mechanics. It duplicates the simplest Boltzmann-Gibbs statistics and must be excluded.
Therefore we conclude that the R\'enyi statistics duplicates the thermodynamic relations
of the familiar Boltzmann-Gibbs statistics in the thermodynamic limit, and leads to
no new result in this respect. This conclusion relies on the particular approach, that the
extra parameter in the entropy formula, $q$ is treated as related to a {\em variable of state},
$z=1/(q-1)$.

%%%%%%%%%%%%%%%%%%%%%%%%%% ACKNOWLEDGMENT %%%%%%%%%%%%
{\bf Acknowledgments:} This work was supported in part by the joint research project of JINR and IFIN-HH, protocol N
3891-3-09/09, the RFBR grant 08-02-01003-a and the MTA-JINR Grant, OTKA K49466, K68108.
We acknowledge fruitful discussions with P.~Levai and V.D.~Toneev.

\end{document}